\documentclass[runningheads]{svmult}

\usepackage{makeidx}   
\usepackage{graphicx}  
\usepackage{subeqnar}  
\usepackage{multicol}  
\usepackage{physprbb}  
\makeindex             

\begin{document}
%
\title*{Prospects for Star Formation Studies with Mid-Infrared Instruments on Large Telescopes}
\toctitle{Prospects for Star Formation Studies with Mid-Infrared Instruments on Large Telescopes}
%
%
\titlerunning{Mid-Infrared Instruments on Large Telescopes}
%
\author{Ray Jayawardhana}
\authorrunning{Ray Jayawardhana}
%
%
\institute{Department of Astronomy, University of California, Berkeley, CA 94720, U.S.A.}

\maketitle              

\begin{abstract}
Imaging and spectroscopic observations in the mid-infrared wavelength range 
(5$\mu$m--30$\mu$m) offer valuable insight into the origins of stars and
planets. Sensitive new array detectors on 8-meter class telescopes make it 
possible to study a wide range of phenomena, from  protoplanetary disks
to starburst galaxies, in unprecedented detail. I review the capablities of 
ground-based mid-infrared instruments (e.g., high spatial resolution) and 
their limitations (e.g., poor sensitivity, small field of view) using 
several examples in the field of star formation, and discuss prospects for 
the near future.
\end{abstract}

\section{Introduction}
The earliest stages of star and planet formation are deeply shrouded in 
cocoons of gas and dust, usually impenetrable at shorter wavelengths. 
Mid-infrared radiation suffers relatively little extinction, and therefore 
provides a valuable probe for investigating these dusty beginnings. The 
5$\mu$m--30$\mu$m range is ideal for observations of the thermal continuum
emitted by warm dust with $T \approx$ 100-300 K, often found in the 
circumstellar environment, and of the spectral features due to a variety of 
atomic (e.g., [Ne II]), molecular (e.g., $H_2$, PAHs) and solid-state (e.g., 
silicates) species.

\section{Promise}
The great promise of ground-based mid-infrared instruments is their high 
spatial resolution. Since imaging in this wavelength regime is (almost) 
diffraction limited, large ground-based telescopes have a significant 
advantage over satellite observatories such as IRAS and ISO with their small 
primary mirrors. 

\section{Limitations}
However, ground-based mid-infrared observations are challenging, to say the 
least, for two primary reasons:
\begin{itemize}
\item
The atmosphere is only partially clear in the 5$\mu$m--30$\mu$m wavelength
range. Since the atmospheric opacity is mainly due to absorption by water 
vapor and carbon dioxide, transmission significantly improves with dryness 
of the site and altitude.
\item
The background emission is dominant over the astronomical sources and variable
on short timescales, making it necessary to continuously measure and subtract 
the sky. (A 300 K blackbody peaks at 10$\mu$m.) Integration times are very 
short (tens of milliseconds), and the use of chopping and nodding is essential.
\end{itemize}

As a result, observations are usually photon-noise limited and the sensitivity 
is relatively poor. Even on an 8-meter telescope, it is difficult to detect
a point source fainter than 1 mJy (or mag$\approx$11.5) in the broad N-band 
centered at 10$\mu$m in a reasonable time. Cooled space-borne observatories 
--such as SIRTF-- will have much better sensitivity.

While mid-infrared detectors have vastly improved over the past two decades, 
most arrays in astronomical use to date have been limited to 128$\times$128 
pixels, providing a small field-of-view ($\sim$10''$\times$10'') on 8-meter 
class telescopes. The new mid-infrared instrument on the ESO 3.6-meter
telescope, TIMMI2, marks a step forward with its 320$\times$240 Si:As array
developed by Raytheon.

The other handicap of the current generation of mid-infrared instruments is
their low spectral resolution, typically $R \approx$100 at 10$\mu$m. However,
COMICS on Subaru achieves $R \sim$ 2000, and VISIR on the VLT is designed to 
do long-slit spectroscopy up to $R \sim$ 30,000.

\section{Science: Results and Prospects}

Despite these limitations, many important scientific results have been
obtained in recent years through the use of mid-infrared instruments on the 
ground. The advent of 8-meter class telescopes enhance the science prospects 
for the near future.

\subsection{Compact HII Regions}
Continuum mid-infrared imaging is useful for tracing the distribution of warm 
dust emission within compact HII regions while narrow-band (e.g., [NeII])
imaging can help identify ionizing fronts. For example, Smith et al. 
(2000) have studied the mid-infrared morphologies of the W49A complex, 
located on the far side of the Galaxy and thus behind some $\approx$300 mag of 
extinction along the line of sight. Their images have sufficient resolution
to investigate the nature of the dust emission from individual sources and
make comparisons with the radio observations. As another recent example,
Chini et al. (2001) produced a 20$\mu$m/10$\mu$m ``temperature map'' of the 
dust in the Orion BN/KL complex, and identified the main sources of heating. 

\subsection{Embedded Protostars}

\begin{figure}[t]
\begin{center}
\includegraphics[width=1.0\textwidth]{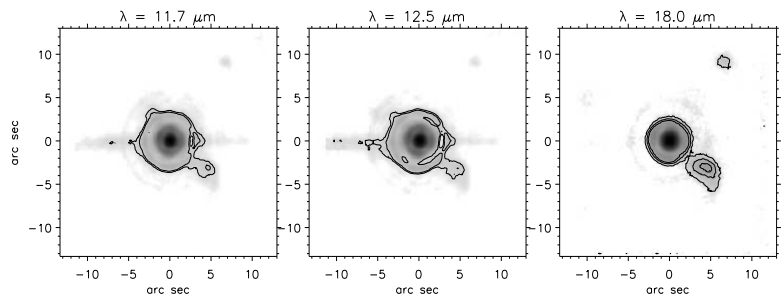}
\end{center}
\caption[]{IRTF/MIRAC3 images of AFGL 2591 at (a) 11.7$\mu$m, (b) 12.5$\mu$m, 
and (c) 18.0$\mu$m. Contours are plotted for 10 and 15 $\sigma$ levels in
panels (a) and (b), and for 5, 10 and 15 $\sigma$ in panel (c). From Marengo
et al. (2000).}
\label{eps1}
\end{figure}

One of the difficulties of probing the formation of protostars is a deficit 
of high spatial resolution observations. Large extinction ($A_{v} \gg$ 10 mag) 
usually prevents optical imaging while millimeter observations typically
have larger beam sizes. Mid-infrared imaging and spectroscopy can help.

For example, Marengo et al. (2000) have recently presented 
sub-arcsecond-resolution images of the high-luminosity young stellar object 
AFGL 2591 and its circumstellar enviornment. Their images, at 11.7, 12.5 
and 18.0 $\mu$m, reveal a knot of emission $\approx$6'' SW of the star, which 
may be evidence for a recent ejection event or an embedded companion star 
(Fig. 1). This knot 
is roughly coincident with a previously seen near-infrared reflection nebula 
and a radio source, and lies within the known large-scale CO outflow. Marengo
et al. also find a new faint NW source which may be another embedded 
lower-luminosity star. The {\it IRAS} mid-infrared spectrum of AFGL 2591 
shows a large silicate absorption feature at 10$\mu$m, implying that the 
primary source is surrounded by an optically thick dusty envelope. 

\subsection{Herbig Ae/Be Stars}
First identified by Herbig (1960) as higher-mass counterparts to young T Tauri
stars, these objects show many of the same signs of activity such as emission 
lines and large infrared excesses. Over the years, several authors have 
attempted to model their spectral energy distributions as dusty envelopes
and/or disks (e.g., Miroshnichenko et al. 1999 and references therein). Recent
mid-infrared imaging by Polomski (2001; Fig. 2) and others reveal companions 
and complex circumstellar environments around many of these sources. 

\begin{figure}[t]
\begin{center}
\includegraphics[width=0.6\textwidth]{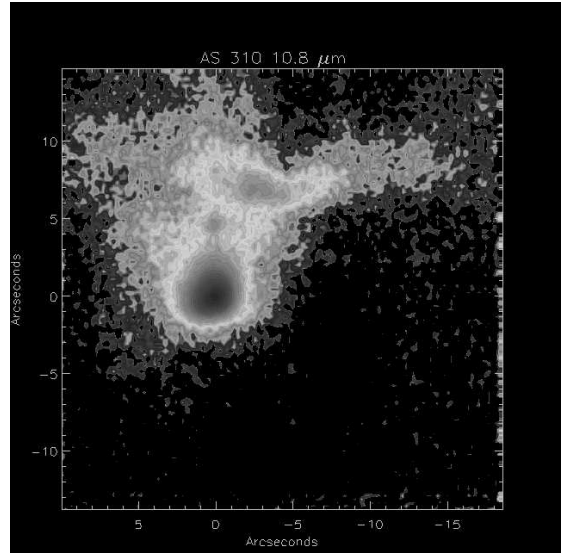}
\end{center}
\caption[]{IRTF/OSCIR image of AS 310 at 10.8$\mu$m, showing complex
circumstellar structure. From Polomski (2001).}
\label{eps2}
\end{figure}

\begin{figure}[t]
\begin{center}
\includegraphics[width=0.6\textwidth]{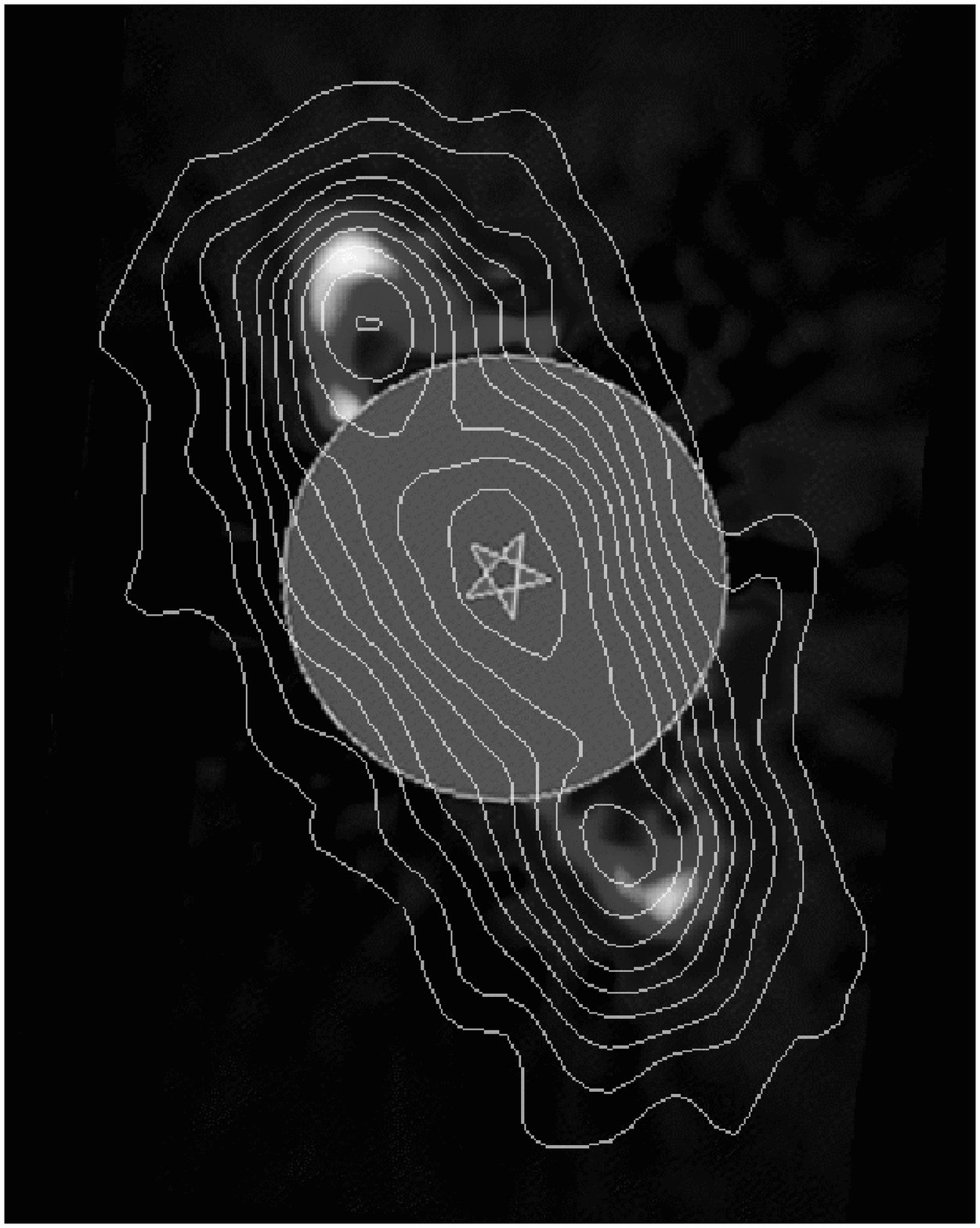}
\end{center}
\caption[]{Overlay of Keck/OSCIR 18.2$\mu$m contours on the 1.1$\mu$m 
HST/NICMOS coronagraphic image of HR 4796A disk. From Telesco et al. (2000).}
\label{eps3}
\end{figure}

\subsection{Circumstellar Disks}
The recent discovery of a spatially-resolved disk around the nearby 
10-Myr-old star HR 4796A is a spectacular demonstration of the power of
high-resolution mid-infared imaging~(Jayawardhana et al. 1998; Koerner et
al. 1998). The surface brightness distribution of the disk is consistent with 
the presence of an inner disk hole of $\sim$50 AU radius, as was first 
suggested by Jura et al. (1993) based on the infrared spectrum, and provides
an important constraint on inner disk evolution timescales. Follow-up
10$\mu$m and 18$\mu$m observations on Keck also revealed tentative evidence
for a brightness asymmetry in the disk (Telesco et al. 2000; Fig. 3) which
may be the result of a forced eccentricity on dust particle orbits by a
companion. 

Mid-infrared spectroscopy, even at a relatively low resolution of 
$R \approx$100, is a useful probe of disk mineralogy. In particular,
the 10$\mu$m silicate feature of some disks appears to be rather similar
to that of comets in the solar system (e.g., Knacke et al. 1993). There
is also some evidence, albeit limited to small samples, for evolution
of the silicate feature over timescales of several million years (e.g., 
Sitko et al. 2000). 

One of the exciting prospects for the near future is nulling interferometry
at 10$\mu$m on the Keck Interferometer and VLTI, with the possibility of
detecting (somewhat higher-surface-brightness) counterparts of the zodiacal 
cloud around nearby stars. Hinz et al. (1998) have already demonstrated 
nulling on an astronomical target with two segments of the MMT. 

\subsection{Starburst Galaxies} 
Observations in the thermal infrared have the potential to identify the
luminosity source in so-called ``ultra-luminous infrared galaxies'' 
determining the size of the emitting region. In many cases, the sources
are unresolved, rather than extended, implying that they are either very 
compact starbursts or dust-enshrouded active galactic nuclei (e.g., Soifer 
et al. 2000). In some starburst galaxies, such as NGC 7469, there is evidence
for a starbursting ring surrounding an AGN (Jayawardhana et al. 1997) and in
the case of NGC 253, a ``super star cluster'' (Keto et al. 1999).

%

\end{document}